\documentclass[10pt,a4paper,twoside=semi,twocolumn,mpinclude=false,footinclude=false]{scrartcl}
\usepackage[left=1.5cm,right=2.2cm,top=3.4cm,bottom=3cm]{geometry}
\usepackage[mono=false]{libertine}
\usepackage[utf8]{inputenc}
\usepackage{amstext}
\usepackage{amsmath}
\usepackage{bbm}
\usepackage{theorem}
\usepackage{graphicx}
\usepackage{tikz}
\usepackage{pgfbaseplot}
\usepackage[backend=biber,style=numeric,sorting=nyt,maxnames=4]{biblatex}
\usepackage{fancyhdr}
\usepackage[noend]{algpseudocode}
\usepackage{xparse}
\usepackage{xcolor}
\usepackage{colorschemes}
\usepackage{xspace}
\usepackage{microtype}
\usepackage{booktabs}
\usepackage[many]{tcolorbox}
\usepackage{varwidth}
\usepackage[final,color,notref,notcite]{showkeys}
\usepackage{inconsolata}
\usepackage[switch]{lineno}
\usepackage[nomessages]{fp}
\usepackage{tocbasic}

\DeclareFieldFormat
  [article,inbook,incollection,inproceedings,patent,thesis,unpublished,misc]
  {title}{\mkbibquote{{\textit{#1}}}}

\colorlet{algotitlebg}{Aluminium3}
\colorlet{algotitlefg}{black}
\colorlet{algocontentbg}{Aluminium1!10!white}
\colorlet{algocontentfg}{black}
\colorlet{algoframe}{Aluminium3}
\colorlet{spotcolor}{SkyBlue3!75!black}
\newcommand{\spotcolor}{\color{spotcolor}}

\addtokomafont{title}{\sffamily\spotcolor}
\addtokomafont{paragraph}{\spotcolor}
\addtokomafont{section}{\spotcolor}
\addtokomafont{subsection}{\spotcolor}
\addtokomafont{subsubsection}{\spotcolor}
\addtokomafont{descriptionlabel}{\spotcolor}
\setkomafont{caption}{}
\setkomafont{captionlabel}{}
\KOMAoptions{numbers=noenddot}

\usetikzlibrary{plotmarks}
\usetikzlibrary{arrows}
\usetikzlibrary{positioning}
\usetikzlibrary{trees}

\tikzset{task/.style={
    rectangle,
    draw=Aluminium4,
    top color=Aluminium1!25!white,
    bottom color=Aluminium1,
    font=\footnotesize,
    inner sep=2pt,
    outer sep=1pt,
    line width=.1pt}}

\colorlet{refkey}{Aluminium4}
\colorlet{labelkey}{ScarletRed3}

\renewcommand{\epsilon}{\varepsilon}
\renewcommand{\phi}{\varphi}

\NewDocumentCommand{\mc}{m}{\ensuremath{\mathcal{#1}}\xspace}
\NewDocumentCommand{\mcH}{}{\ensuremath{\mc{H}}\xspace}

\NewDocumentCommand{\mcO}{}{\ensuremath{\mc{O}}\xspace}
\NewDocumentCommand{\mcP}{}{\ensuremath{\mc{P}}\xspace}
\NewDocumentCommand{\mcL}{}{\ensuremath{\mc{L}}\xspace}

\NewDocumentCommand{\mcA}{}{\ensuremath{\mc{A}}\xspace}

\NewDocumentCommand{\landau}{m}{\ensuremath{\mcO\left(#1\right)}\xspace}
\NewDocumentCommand{\set}{m}{\ensuremath{\left\{ #1 \right\}}\xspace}

\NewDocumentCommand{\rank}{}{\ensuremath{\operatorname{rank}}\xspace}

\NewDocumentCommand{\N}{}{\ensuremath{\mathbbm{N}}}
\NewDocumentCommand{\R}{}{\ensuremath{\mathbbm{R}}}

\NewDocumentCommand{\K}{}{\ensuremath{\mathbbm{K}}}

\NewDocumentCommand{\op}{m}{\ensuremath{\operatorname{#1}}\xspace}

\NewDocumentCommand{\troot}{m}{\ensuremath{\op{root}(#1)}\xspace}
\NewDocumentCommand{\sons}{m}{\ensuremath{\op{sons}(#1)}\xspace}
\NewDocumentCommand{\function}{ m o }{\textnormal{{\ttfamily\textsc{#1}\IfNoValueF{#2}{(#2)}}}\xspace}
\newtcolorbox[auto counter]{algorithmbox}[2][]{  %% ,number within=section
  floatplacement=htb,
  float,
  enhanced,
  size=fbox,
  before skip=3mm,
  colbacktitle=algotitlebg!25!white,
  title style={top color=algotitlebg!20!white,bottom color=algotitlebg!30!white},
  colback=algocontentbg,
  colframe=algoframe,
  arc=1pt,
  halign=flush left,
  coltitle=algotitlefg,
  title={\textbf{Algorithm~\thetcbcounter:} #2},
  label={#1}}
\newtcolorbox[auto counter]{inlinealgorithmbox}{ %% ,number within=section
  size=tight,
  boxsep=0.5mm,
  before skip=2mm,
  after skip=2mm,
  colback=algocontentbg,
  colframe=algoframe,
  arc=1pt,
  halign=flush left}
\newtcolorbox[auto counter]{figurealgorithmbox}{ %% ,number within=section
  size=tight,
  before skip=0mm,
  colback=algocontentbg,
  boxrule=0pt,
  colframe=white,
  frame hidden,
  halign=flush left}
\NewDocumentEnvironment{inlinealgorithm}%
                       {}%
                       {\begin{inlinealgorithmbox}
                           \hspace{-0.9em}\begin{varwidth}{\linewidth}\ttfamily\small%
                           \begin{algorithmic}}%
                       {\end{algorithmic}\end{varwidth}\end{inlinealgorithmbox}}

\DeclareNewTOC[
  type = fltalgorithm ,
  float ,
]{fltalgorithm}
                 
\NewDocumentEnvironment{algorithm}%
                       {m m}%
                       {\begin{algorithmbox}[#2]{#1}%
                           \hspace{-1em}\begin{varwidth}{\linewidth}\ttfamily\small%
                           \begin{algorithmic}}%
                       {\end{algorithmic}\end{varwidth}\end{algorithmbox}}
\NewDocumentEnvironment{algorithm*}%
                       {m m}%
                       {\begin{fltalgorithm*}\begin{algorithmbox}[#2]{#1}%
                           \hspace{-1em}\begin{varwidth}{\linewidth}\ttfamily\small%
                           \begin{algorithmic}}%
                       {\end{algorithmic}\end{varwidth}\end{algorithmbox}\end{fltalgorithm*}}

\NewDocumentEnvironment{figurealgorithm}%
                       {}%
                       {\begin{figurealgorithmbox}
                           \hspace{-.75em}\begin{varwidth}{\linewidth}\ttfamily\small%
                           \begin{algorithmic}}%
                       {\end{algorithmic}\end{varwidth}\end{figurealgorithmbox}}

\algnewcommand{\SFor}[2]{\State \algorithmicfor\ #1\ \algorithmicdo\ #2\ }
\algnewcommand{\SForAll}[2]{\State \algorithmicforall\ #1\ \algorithmicdo\ #2\ }
\algnewcommand{\SIf}[2]{\State \algorithmicif\ #1\ \algorithmicthen\ #2\ }
\algnewcommand{\SElsIf}[2]{\State \algorithmicelse\ \algorithmicif\ #1\ \algorithmicthen\ #2\ }
\algnewcommand{\SElse}[1]{\State \algorithmicelse\ #1\ }

\usepgflibrary{fpu}

\theoremstyle{plain}
\theoremheaderfont{\bfseries \upshape}
\newtheorem{define}{Definition}[section]
\newtheorem{remark}[define]{Remark}

\setkomafont{author}{\normalsize}
\setkomafont{date}{\normalsize}

\makeatletter
\renewcommand{\maketitle}{
  \parbox{0.96\linewidth}{\centering{\usekomafont{title}{\huge \@title\par}}}
  \vskip 1.0em
  \parbox{0.96\linewidth}{\centering{\usekomafont{title}{\large \@subtitle\par}}}
  \vskip 1.5em
  \parbox{0.96\linewidth}{%
    \centering
    \usekomafont{author}{%
      \lineskip 0.75em
      \begin{tabular}[t]{c}
        \@author
      \end{tabular}\par
    }%
  }%
  \vskip 1.5em
  \parbox{0.96\linewidth}{\centering{\usekomafont{date}{\@date\par}}}
  \vskip 2.0em
}
\makeatother

\renewenvironment{abstract}{%
  \centering
  \setlength{\parindent}{0pt}
  \begin{minipage}{\linewidth}%
    \setlength{\parindent}{0pt}%
    \setlength{\parskip}{.5em}%
    \textbf{\textcolor{spotcolor}{Abstract}}%
  }{\end{minipage}
}

\NewDocumentCommand{\clt}{}{\ensuremath{\tau}\xspace}
\NewDocumentCommand{\cls}{}{\ensuremath{\sigma}\xspace}
\NewDocumentCommand{\clr}{}{\ensuremath{\rho}\xspace}

\NewDocumentCommand{\prnfp}{ m m }{{\ttfamily{1}-{#1}-{#2}}\xspace}

\begin{document}

\title{Hierarchical Lowrank Arithmetic with Binary Compression}
\author{{\large Ronald Kriemann}\small\\
  MPI for Mathematics i.t.S.\\
  Leipzig, Germany\\
  rok@mis.mpg.de
}
\maketitle

\begin{abstract}
  With lowrank approximation the storage requirements for dense data are reduced down to linear complexity and
  with the addition of hierarchy this also works for data without global lowrank properties. However, the lowrank
  factors itself are often still stored using double precision numbers. Newer approaches exploit the different
  IEEE754 floating point formats available nowadays in a mixed precision approach. However, these formats show
  a significant gap in storage (and accuracy), e.g. between half, single and double precision. We therefore
  look beyond these standard formats and use adaptive compression for storing the lowrank and dense data and
  investigate how that affects the arithmetic of such matrices.
  

\end{abstract}

\pagestyle{fancy}
\thispagestyle{plain}
\fancyhf{}
\fancyhf[HLE]{\footnotesize\thepage\hfill R. Kriemann}
\fancyhf[HRO]{\footnotesize \mcH-Arithmetic with Binary Compression\hfill\thepage}
\fancyhf[FC]{}
\renewcommand{\headrulewidth}{0.4pt}
\renewcommand{\footrulewidth}{0pt}
\thispagestyle{empty}

%%%%%%%%%%%%%%%%%%%%%%%%%%%%%%%%%%%%%%%%%%%%%%%%%%%%%%%%%%%%%%%%%%%%%%%%%%%%%%%%%%%%%%%%%
%%%%%%%%%%%%%%%%%%%%%%%%%%%%%%%%%%%%%%%%%%%%%%%%%%%%%%%%%%%%%%%%%%%%%%%%%%%%%%%%%%%%%%%%%

%% \linenumbers

%%%%%%%%%%%%%%%%%%%%%%%%%%%%%%%%%%%%%%%%%%%%%%%%%%%%%%%%%%%%%%%%%%%%%%%%%%%%%%%%%%%
%%%%%%%%%%%%%%%%%%%%%%%%%%%%%%%%%%%%%%%%%%%%%%%%%%%%%%%%%%%%%%%%%%%%%%%%%%%%%%%%%%%
%%%%%%%%%%%%%%%%%%%%%%%%%%%%%%%%%%%%%%%%%%%%%%%%%%%%%%%%%%%%%%%%%%%%%%%%%%%%%%%%%%%

\section{Introduction} \label{sec:intro}

In the last decades lowrank techniques proved very effective for representing dense data with (almost) optimal storage
complexity. Especially in the form of hierarchical matrices (\mcH-matrices), as first introduced in
\cite{Hackbusch:1999}, it allows to handle operators of much larger problem sizes thanks to near linear storage and
arithmetic complexity. A large variety of different applications demonstrate the huge advantage of lowrank storage.

Though the complexity reduction is the major contributor for this success, there is still room for improvements,
in particular in the raw storage of floating point numbers for the dense and lowrank blocks within an
\mcH-matrix. In practice, these blocks are often stored in double precision format (FP64). One of the reasons being the
approximation algorithms used for lowrank arithmetic, e.g., singular value decomposition, which have more strict
accuracy requirements sometimes surpassing even the capabilities of single precision (FP32) arithmetic. However, the
machine precision or \emph{unit roundoff} associated to this is typically much smaller than the already introduced
lowrank approximation error.

The IEEE-754 standard, which defines both storage formats, also provides further floating point schemes, some of which
are presented in Table~\ref{tab:ieee754}. Especially in the recent years, half precision formats, e.g., FP16, BF16 or
TF32, have been used on GPUs. However, all these only provide a fixed arithmetic accuracy. Compression schemes for
\mcH-matrices however, should support a variable precision as the lowrank approximation accuracy may change in a wide
range depending on the application.

\begin{table}[tbp]
  \centering
  \begin{tabular}{llll}
    & \multicolumn{1}{c}{\textbf{1-\(\mathbf{e}\)-\(\mathbf{m}\)}} 
    & \multicolumn{1}{c}{\textbf{Bits}}
    & \multicolumn{1}{c}{\textbf{Unit Roundoff}} % 2^{-(m+1)} due to implicit bit in exponent
    \\
    \toprule
    FP64 & \prnfp{11}{52} & 64 & \(1.1 \times 10^{-16}\) \\
    FP32 & \prnfp{8}{23}  & 32 & \(6.0 \times 10^{-8}\)  \\
    TF32 & \prnfp{8}{10}  & 19 & \(4.9 \times 10^{-4}\) \\
    BF16 & \prnfp{8}{7}   & 16 & \(3.9 \times 10^{-3}\) \\
    FP16 & \prnfp{5}{10}  & 16 & \(4.9 \times 10^{-4}\) \\
    FP8  & \prnfp{4}{3}   &  8 & \(6.3 \times 10^{-2}\)
  \end{tabular}
  \caption{Floating point formats based on the IEEE-754 standard}
  \label{tab:ieee754}
\end{table}

An alternative are \emph{mixed precision} approaches. In \cite{AbdCaoPeiBos:2022} single or half precision
representation was chosen for a full lowrank block depending on its Frobenius norm compared to the norm of the
matrix. However, this still requires that a lowrank block can be represented in the corresponding floating point
format.

In contrast to this, in \cite{AmeBoiBut:2022} the lowrank factors \(U \in \R^{n \times k}\) and \(V \in \R^{m \times
  k}\) are decomposed into subblocks \(U = (U_1,U_2,\ldots), U_i \in \R^{n \times k_i}\) and \(V = (V_1,V_2,\ldots), V_i
\in \R^{m \times k_i}\) with the size \(k_i\) of each subblock depending on the singular values of \(U V^H\) and the
unit roundoff of the floating point format used to represent \(U_i,V_i\). With this, even for a high accuracy,
low precision formats can be used.

A different approach was implemented in \cite{KriLtaLuo:2022} by using the floating point compression library ZFP
\cite{Lindstrom:2014} to further compress the lowrank and dense data blocks with an adaptive accuracy based on the user
defined lowrank approximation precision. With this, the overall memory consumption could further be reduced
significantly.

In principle, other floating point compression schemes could also be applied to \mcH-matrices, e.g., SZ
\cite{DiCap:2016,LiDi:2018} (or SZ3 \cite{ZhaDiDmi:2021}), which demonstrates a very good compression rate for a wide
range of scientific data. A different approach to floating point compression is implemented in MGARD \cite{mgard} which
is based on multigrid techniques.

Related to lowrank compression are methods based on Tucker decomposition, e.g., TuckerMPI \cite{TuckerMPI}, TTHRESH
\cite{tthresh} or ATC \cite{atc}, as it can be considered the generalization of the same concept to higher
dimensions. TTHRESH and ATC also use \emph{bitplane truncation} to produce a minimal memory representation of the
floating point data. However, these compression techniques are only of limited use in the context of \mcH-matrices with
its two-dimensional data blocks for which Tucker decomposition corresponds to standard SVD.

Another alternative is lossless compression, e.g. as implemented in the widely used libraries Zstd \cite{Zstd}, LZ4
\cite{LZ4} or zlib \cite{zlib}. However, experiments with \mcH-matrices yielded only a reduction of the memory size by
about 10\% at best.

When extending the compression of lowrank or dense data within an \mcH-matrix to the full \mcH-matrix arithmetic, it is
not easily possible to perform arithmetic operations directly with the compressed data, especially as lowrank arithmetic
relies on more complex functions like SVD or QR factorization. Instead, one can perform computations still in double (or
single) precision and only store values in a compressed form. This \emph{approximate storage} concept is introduced in
\cite{AnzGruQui:2019} for sparse matrix arithmetic.

In this work, we apply it to full \mcH-matrix arithmetic, i.e., matrix-vector multiplication and \mcH-LU
factorization. While the former only requires decompression of data, LU factorization includes updates to matrix blocks
involving matrix multiplication and lowrank truncation during which constant decompression and recompression of the
matrix data is performed. We will comare different compression schemes in terms of compression ratio and arithmetic
performance.

For algorithms limited by the memory bandwidth, a possible side effect of the reduced storage size is an increase in
performance since less data needs to be loaded or stored. Though this is not the main focus of this work, some examples
of this effect are visible also for \mcH-matrix algorithms.

The rest of this work is structured as follows: in Section~\ref{sec:hmat} basic definitions and algorithms for
\mcH-matrices are introduced. Section~\ref{sec:zhmat} will discuss compression of the data blocks within \mcH-matrices
and Section~\ref{sec:zarith} the special properties of compressed \mcH-arithmetic. Numerical experiments will be
presented in Section~\ref{sec:numexp}, followed by a conclusion in Section~\ref{sec:conclude}.

%%%%%%%%%%%%%%%%%%%%%%%%%%%%%%%%%%%%%%%%%%%%%%%%%%%%%%%%%%%%%%%%%%%%%%%%%%%%%%%%%%%
%%%%%%%%%%%%%%%%%%%%%%%%%%%%%%%%%%%%%%%%%%%%%%%%%%%%%%%%%%%%%%%%%%%%%%%%%%%%%%%%%%%
%%%%%%%%%%%%%%%%%%%%%%%%%%%%%%%%%%%%%%%%%%%%%%%%%%%%%%%%%%%%%%%%%%%%%%%%%%%%%%%%%%%

\section{\mcH-Matrices and \mcH-Arithmetic} \label{sec:hmat}

%%%%%%%%%%%%%%%%%%%%%%%%%%%%%%%%%%%%%%%%%%%%%%%%%%%%%%%%%%%%%%%%%%%%%%%%%%%%%%%%%%%

\subsection{Definitions} \label{sec:hdef}

For an indexset \(I\) we define the \emph{cluster tree} (or \mcH-tree) as the hierarchical partitioning of \(I\) into
disjoint sub-sets of \(I\):
\begin{define}[Cluster Tree]
  Let \(T_I = (V,E)\) be a tree with \(V \subset \mathcal{P}(I)\). \(T_I\) is called a \emph{cluster tree} over \(I\) if
  \begin{enumerate}
  \item \(I = \troot{T_I}\) and
  \item for all \(v \in V\) with \(\sons{v} \ne \emptyset : v = \dot\cup_{v' \in \sons{v}} v'\).
  \end{enumerate}
  A node in \(T_I\) is also called a \emph{cluster} and we write \(\clt \in T_I\) if \(\clt \in V\). The set of leaves
  of \(T_I\) is denoted by \(\mcL(T_I)\).
\end{define}

Similar to a cluster tree we can extend the hierarchical partitioning to the product \(I \times J\) of two index sets
\(I, J\), while restricting the possible set of nodes by given cluster trees \(T_I\) and \(T_J\) over \(I\) and \(J\),
respectively. Furthermore, the set of leaves will be defined by an \emph{admissibility condition}. In the literature,
various examples of admissibility can found, e.g. standard \cite{HackKhor:2000}, weak \cite{HackKhorKrie:2004} or
off-diagonal admissibility \cite{ChaDew:2005,AmbDar:2013}.

\begin{define}[Block Tree]
  Let \(T_I, T_J\) be two cluster trees and let \(\op{adm} : T_I \times T_J \to \mathbbm{B} \). The \emph{block tree}
  \(T = T_{I \times J} \) is recursively defined starting with \( \troot{T} = (I,J) \):
  % \begin{itemize}
  %   \item \(\sons{\clt,\cls} = \emptyset\) if \(\op{adm}(\clt \times \cls) = \) true,
  %   \item \(\sons{\clt,\cls} = \emptyset\) if \(\sons{\clt} = \emptyset\) or \(\sons{\cls} = \emptyset\),
  %   \item \(\set{ (\clt',\cls') \,:\, t' \in \sons{\clt}, s' \in \sons{\cls} }\) otherwise.
  % \end{itemize}
  \begin{align*}
    & \sons{\clt,\cls} = \\
    & \begin{cases} \emptyset, \textnormal{ if } \op{adm}(\clt,\cls) = \textnormal{true} \vee
      \sons{\clt} = \emptyset \vee \sons{\cls} = \emptyset,\\
      \set{ (\clt',\cls') \,:\, \clt' \in \sons{\clt}, \cls' \in \sons{\cls} } \textnormal{ else}.
    \end{cases}
  \end{align*}
  A node in \(T\) is also called a \emph{block}. Again, the set of leaves of \(T\) is denoted by \(\mcL(T) := \set{ b
    \in T \;:\; \sons{b} = \emptyset}\). By \(\mcL_{lr}(T) = \set{ b \in \mcL \;:\; \op{adm}(b) = \textnormal{true} }\)
  the set of admissible leaves is denoted.
\end{define}

The admissibility condition is used to detect blocks in $T$ which can be efficiently approximated by lowrank matrices
with a predefined rank \(k\), i.e., blocks \(b\) with \(\op{adm}(b) = \textnormal{true}\). The set of all such matrices
forms the set of \mcH-matrices:
\begin{define}[\mcH-Matrix]
  For a block tree $T$ over cluster trees $T_I, T_J$ and $k \in \N$, the set of \mcH-matrices $\mcH(T,k)$ is defined as
  \begin{align*}
    \mcH(T,k) := \{& M \in \R^{I \times J} \;:\; \forall (\clt,\cls) \in \mcL(T) : \\
                   & \rank(M_{\clt,\cls}) \le k \vee \clt \in \mcL(T_I) \vee \cls \in \mcL(T_J)\}
  \end{align*}
  Here, \(M_{\clt,\cls}\) refers to the sub-block \(M|_{\clt \times \cls}\).
\end{define}

In practice the constant rank \(k\) is typically replaced by a fixed lowrank approximation accuracy \(\varepsilon > 0\)
as the resulting \mcH-matrices are often more memory efficient. For this we assume for an admissible block
\(M_{\clt,\cls}\):
\begin{equation} \label{eqn:epsilon}
||M_{\clt,\cls} - U_{\clt,\cls} V_{\clt,\cls}^H|| \le \varepsilon ||M_{\clt,\cls}|| \ .
\end{equation}

%%%%%%%%%%%%%%%%%%%%%%%%%%%%%%%%%%%%%%%%%%%%%%%%%%%%%%%%%%%%%%%%%%%%%%%%%%%%%%%%%%%

\subsection{\mcH-Arithmetic} \label{sec:harith}

During arithmetic, the most important operation is lowrank truncation as it is required after every update to a lowrank
block and also normally more costly than other operations, e.g., dense matrix addition or multiplication.

In the literature different forms of lowrank truncation exist, with the SVD being the classical form
\cite{GrasedyckHackbusch:2003} and shown in Algorithm~\ref{alg:lrsvd}. There, the function \function{rank} determines
the truncation rank based on the singular values in \(S_s\) and maybe application dependent.

\begin{algorithm}{Lowrank Truncation via SVD}{alg:lrsvd}
  \Procedure{truncate}{in: $U, V$, out: $W, X$}
  \State \([Q_U, R_U] := \function{qr}[\(U\)]\);
  \State \([Q_V, R_V] := \function{qr}[\(V\)]\);
  \State \([U_s,S_s,V_s] := \function{svd}[\(R_U \cdot R_V'\)]\);
  \State \(k := \function{rank}[ \(S_s\) ];\)
  \State \(W := Q_U \cdot  U_s(1:k,:) \cdot S_s(1:k,1:k) \);
  \State \(X := Q_V \cdot V_s(1:k,:)\);
  \EndProcedure
\end{algorithm}

Other truncation algorithms use rank revealing QR or randomized algorithms\cite{HalMarTro:2011}.

Arithmetic functions for \mcH-matrices are typically formulated in a recursive way. For the matrix multiplication 
\(C := \alpha A \cdot B + C\) with \(A, B, C \in \mcH(T_{I \times I},k)\) such a formulation can be found in
Algorithm~\ref{alg:hmul} where the cluster tree \(T(I)\) is assumed to be binary.

\begin{algorithm}{\mcH-Matrix Multiplication}{alg:hmul}
  \Procedure{hmul}{$\alpha,A_{\clt,\rho}, B_{\rho,\cls},C_{\clt,\cls}$}
  \If{\(\set{(\clt,\rho),(\rho,\cls),(\clt,\cls)} \cap \mcL(T) = \emptyset \)}
  \For{\(i,j,\ell \in \set{0,1}\)}
  \State \function{hmul}[\(\alpha,A_{\clt_i,\rho_\ell},B_{\rho_\ell,\cls_j},C_{\clt_i,\cls_j}\)];
  \EndFor
  \Else
  \State \(C_{\clt,\cls} := C_{\clt,\cls} + \alpha A_{\clt,\rho} B_{\rho,\cls}\); \Comment{Block Update}
  \EndIf
  \EndProcedure
\end{algorithm}

The last line of Algorithm~\ref{alg:hmul} forms the actual update of the corresponding matrix block and normally
involves a lowrank truncation if \(C_{\clt,\cls}\) is a lowrank matrix block.

Analog recursive functions can also be formulated for triangular matrix solves or the LU factorization. 

An alternative formulation of the \mcH-matrix arithmetic is described in \cite{Bor:2019}. There, instead of applying all
updates to lowrank blocks directly, these updates are accumulated during the recursion in an extra \emph{accumulator}
matrix \(\mcA_{\clt,\cls} \in \K^{\clt \times \cls}\).
% with splitting and shifting down this matrix in each recursion step.
Furthermore, non-computable, so-called \emph{pending} updates, e.g., involving further recursion, are collected in
sets \(\mcP_{\clt,\cls} \subseteq \set{ (A_{\clt,\clr},B_{\clr,\clt})\;:\; \set{(\clt,\clr),\clr,\cls)} \subset
  T}\).
% Updates from these sets are also split into subsets during recursion.
Both, \(\mcA_{\clt,\cls}\) and \(\mcP_{\clt,\cls}\), are initialized to zero at the start of the multiplication and as
such don't consume any memory as lowrank matrices are used.

The \mcH-multiplication Algorithm~\ref{alg:hmulaccu} recursively follows the structure of the \mcH-matrix \(C\) from
root to leaves. For each sub-block of \(C_{\clt,\cls}\), first all computable pending updates, i.e., with at least one
dense or lowrank block, are evaluated with the result being applied to the corresponding accumulator matrix. If
\(C_{\clt,\cls}\) has sub-blocks, the remaining pending updates are split into sub-sets for all sub-blocks, which mimics
the standard triple-loop in recursive matrix multiplication in Algorithm~\ref{alg:hmul}. Afterwards, the algorithm
recurses for each sub-block with also the accumulator matrix being restricted to each of these sub-blocks. Finally, if
\(C_{\clt,\cls}\) is a leaf matrix block, all updates accumulated in \(\mcA_{\clt,\cls}\) are applied in a single step.

\begin{algorithm}{Accumulated \mcH-Matrix Multiplication}{alg:hmulaccu}
  \Procedure{hmulaccu}{$C_{\clt,\cls},\mcA_{\clt,\cls}, \mcP_{\clt,\cls}$}
  \ForAll{\((A_{\clt,\clr}, B_{\clr,\cls}) \in \mcP_{\clt,\cls}\)}
    \If{\((\clt,\clr) \in \mcL(T) \textnormal{ or } (\clr,\cls) \in \mcL(T)\)}
      \State \(\mcA_{\clt,\cls} := \mcA_{\clt,\cls} + A_{\clt,\clr} B_{\clr,\cls}\);
      \State \(\mcP_{\clt,\cls} := \mcP_{\clt,\cls} \setminus \set{ ( A_{\clt,\clr}, B_{\clr,\cls} ) }\);
    \EndIf
  \EndFor
  \If{\((\clt,\cls) \not\in \mcL(T)\)}
    \For{\(i,j \in \set{0,1}\)}
      \State \(\mcP_{\clt_i,\cls_j} := \cup_{l=0}^{1} \left\{(A_{\clt,\clr}|_{\clt_i,\clr_{\ell}}, B_{\clr,\cls}|_{\clr_{\ell},\cls_j}) :\right.\)
      \State \hspace{2.25cm}\(\left.(A_{\clt,\clr}, B_{\clr,\cls}) \in \mcP_{\clt,\cls}\right\}\);
      \State \function{hmulaccu}[\(C_{\clt_i,\cls_j}, \mcA_{\clt,\cls}|_{\clt_i,\cls_j},\mcP_{\clt_i,\cls_j}\)];
    \EndFor
  \Else
  \State \(C_{\clt,\cls} := C_{\clt,\cls} + \mcA_{\clt,\cls}\);
  \EndIf
  \EndProcedure
\end{algorithm}

\begin{remark} \label{rem:memaccu}
  Due to the top-down approach for accumulator based \mcH-arithmetic, only non-zero accumulator matrices for the current
  recursion path exist, which bounds its number by \(\landau{\log n}\) for a sequential implementation. In the parallel
  case, this typicall increases linearly with the number of parallel execution paths. The additional memory overhead for the
  accumulator matrices is therefore negligible compared to the memory of the involved \mcH-matrices.
\end{remark}

Independent on the arithmetic version, effective parallelization of the \mcH-arithmetic on shared-memory systems can be
done by identification of the arithmetic tasks, which, in the case of \mcH-matrix multiplication are defined by the last
line in Algorithm~\ref{alg:hmul} and the dependencies between these tasks, which together form a directed acyclic graph
(DAG) and can be executed by a scheduling system. However, while the task definition is straight-forward and can easily
be performed with the recursive \mcH-arithmetic functions, the dependencies are more involved as different, unrelated
parts of the block tree are affected\cite{Kri:2013,BoeChrKri:2022}.

%%%%%%%%%%%%%%%%%%%%%%%%%%%%%%%%%%%%%%%%%%%%%%%%%%%%%%%%%%%%%%%%%%%%%%%%%%%%%%%%%%%
%%%%%%%%%%%%%%%%%%%%%%%%%%%%%%%%%%%%%%%%%%%%%%%%%%%%%%%%%%%%%%%%%%%%%%%%%%%%%%%%%%%
%%%%%%%%%%%%%%%%%%%%%%%%%%%%%%%%%%%%%%%%%%%%%%%%%%%%%%%%%%%%%%%%%%%%%%%%%%%%%%%%%%%

\section{Compressed \mcH-matrices} \label{sec:zhmat}

\mcH-matrices use dense memory blocks for inadmissible sub-blocks and for the lowrank factors \(U,V\). These data blocks
are of interest for floating point compression. This is especially true for the lowrank factors as lowrank approximation
already introduces an error, typically much higher compared to the unit roundoff of the FP64 format.

The storage of these data blocks is considered to be separate even on the block level, i.e., \(U\) and \(V\) are stored
independently from each other, in contrast to a joined storage. This simplifies the data conversion during
\mcH-arithmetic as both factors may be used at different times during arithmetic operations, e.g., \mcH-LU factorization
or matrix solves.

In the following we will use the lowrank approximation accuracy \(\varepsilon\) from \eqref{eqn:epsilon} also for the
compression accuracy, i.e., the same error that was permitted for the lowrank approximation may also be allowed for
floating point compression.

\subsection{Floating Point Compression Libraries} \label{sec:zfpsz}

Floating point compressors like ZFP or SZ/SZ3 seem an obvious choice when the dense data blocks in the \mcH-matrices
should be compressed further, assuming a sufficient error control is provided by the compression schemes.

For ZFP, this was possible with \emph{fixed rate} mode, which uses a constant number of bits for each data
block\footnote{ZFP handles sub-blocks of size \(4^d\) seperately, with \(d\) being the dimension of the data.}. The
(bit-) rate for a given accuracy \(\varepsilon\) was determined by experiments to be \(\lceil|\log_2
\varepsilon|\rceil+2\). Unfortunately, fixed-precision or fixed accuracy modes also provided by ZFP and potentially
superior to the fixed-rate mode were much more difficult in the context of \mcH-matrices and did not allowed a reliable
error control.

\begin{figure}[htb]
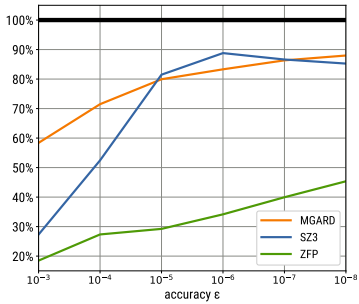

  \centering
  \pgfdeclareimage[width=.28\textwidth]{lapmemsz3mgard}{data/mvm/laplace-sphere-6/zmem-laplace}
  \pgfuseimage{lapmemsz3mgard}
  \caption{Compression rates for SZ3 and MGARD for Laplace model problem \eqref{eqn:slp} with \(n=32.768\).}
  \label{fig:sz3mgard}
\end{figure}

SZ and SZ3 support more options for handling the error, e.g., absolute or relative error bounds, peak
signal to noise ratio or Frobenius error. However, in experiments, difficulties with the compression of \mcH-matrix
data were observed if the accuracy \(\varepsilon\) exceeded \(10^{-5}\), i.e., \(\varepsilon \le 10^{-5}\) (see
Figure~\ref{fig:sz3mgard}). The same holds for MGARD.

Though the cause of these problems is unknown, both SZ/SZ3 and MGARD rely on some form of approximability of the given
data, either by curve fitting (SZ/SZ3) or by multigrid techniques (MGARD). While such properties may be given in the
original dense data, within \mcH-matrices the lowrank factors already represent a form of compressed data which may not
have these approximation properties anymore. Therefore, SZ/SZ3 and MGARD were not considered for further experiments.

\subsection{IEEE-754 based Compression} \label{sec:ieee754}

As already discussed, using a single floating point format will not suffice for storing data within an
\mcH-matrix. Instead, a format with a much higher adaptivity towards lowrank approximation accuracy is needed.

% \begin{figure}[tbp]
%   \centering
%   \pgfdeclareimage[width=.45\textwidth]{memmp}{data/mem-mixedprec}
%   \pgfuseimage{memmp}
%   \caption{Memory usage of mixed precision approach.}
%   \label{fig:memmp}
% \end{figure}

Discussed in \cite{AnzGruQui:2019} is a format for a fully adaptive choice of the mantissa and exponent bits in the
IEEE-754 scheme, depending on the floating point values in the individual blocks of the block-Jacobi preconditioner. The
dense and lowrank blocks of \mcH-matrices provide also a partitioning of the matrix data and by that also permit an per
sub-block adaptive choice of the storage layout. FlexFloat \cite{FlexFloat} implements the same adaptive choice of
mantissa and exponent bits. However, the in-memory storage of FlexFloat does not make use of it and as such, no actual
memory savings can be expected.

While the number of mantissa bits depends on the accuracy requirements of the lowrank approximation, the exponent size
depends on the \emph{dynamic range} of the data values, i.e., the base 10 logarithm of the ratio between the largest and
smallest (absolute) value. For the standard floating point formats the corresponding dynamic range is typically huge, as
can be seen in Table~\ref{tab:dynrange}.

\begin{table}[htb]
  \centering
  \begin{tabular}{cccccc}
    \textbf{FP64} & \textbf{FP32} & \textbf{TF32} & \textbf{BF16} & \textbf{FP16} & \textbf{FP8} \\
    \hline
    631 & 83 & 79 & 78 & 12 & 5 \\
  \end{tabular}
  \caption{Dynamic range of standard floating point formats.}
  \label{tab:dynrange}
\end{table}

\begin{figure}[htb]
  \centering
  \pgfdeclareimage[width=.22\textwidth]{laplacedynrange}{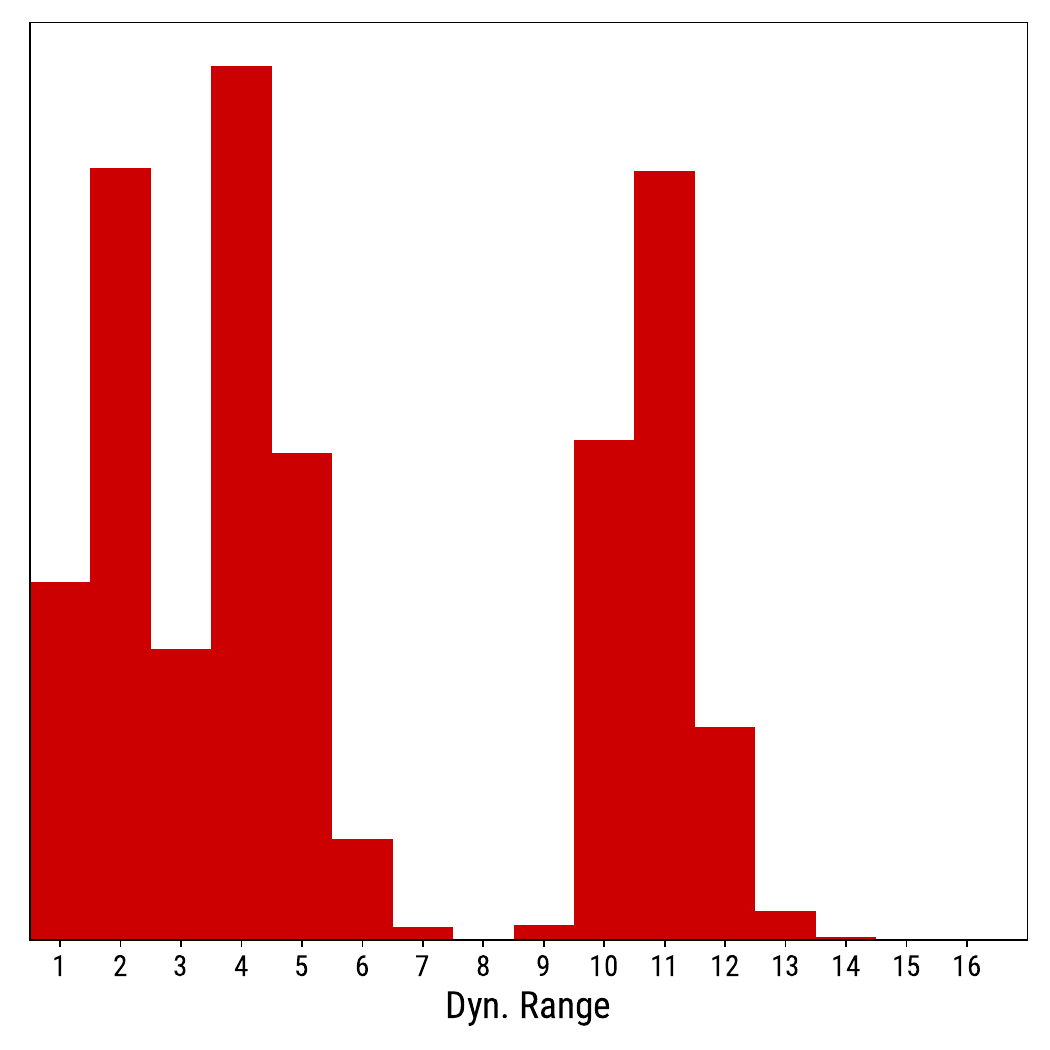}
  \pgfdeclareimage[width=.22\textwidth]{materndynrange}{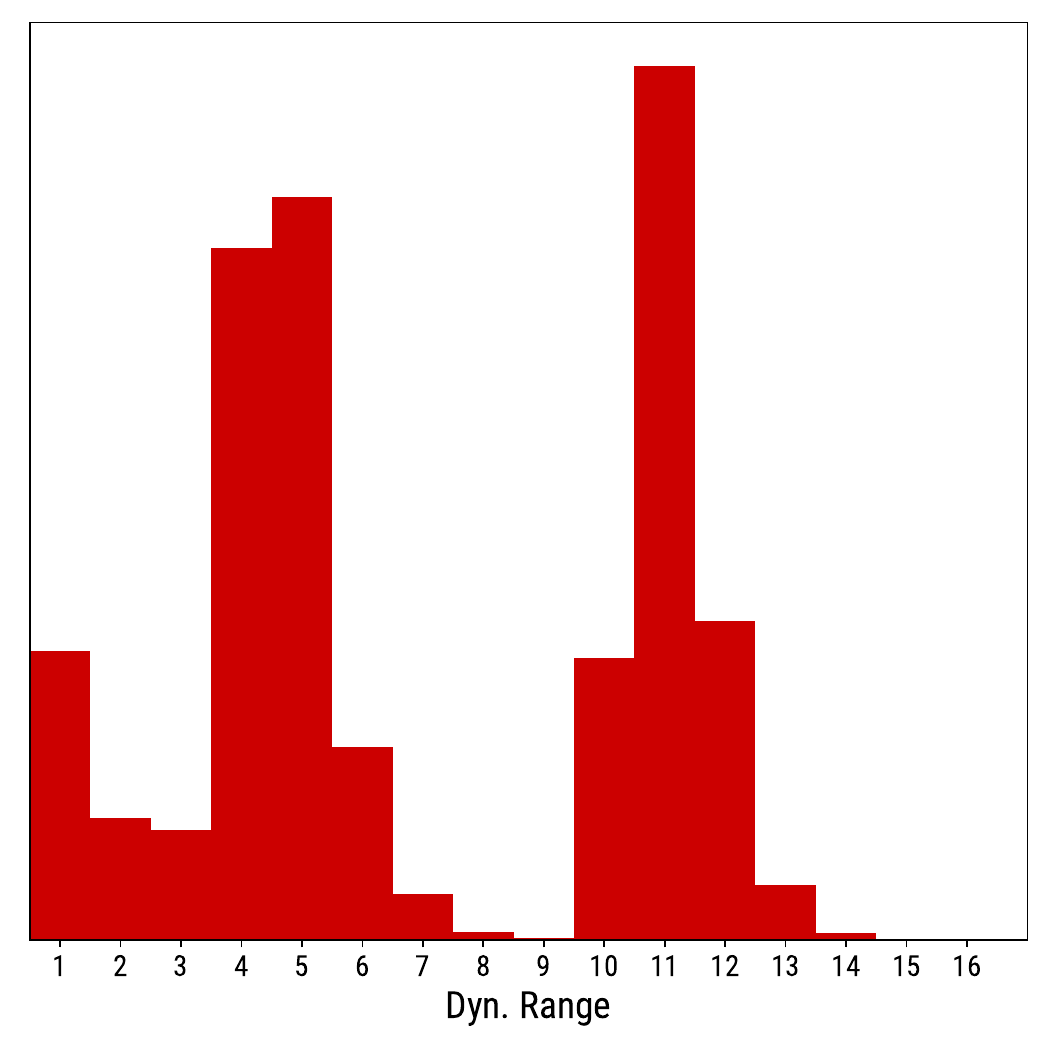}
  \begin{tabular}{cc}
    Laplace SLP & Mat\'ern covariance \\ 
    \pgfuseimage{laplacedynrange} &
    \pgfuseimage{materndynrange}
  \end{tabular}
  \caption{Dynamic range of per data block floating point values during \mcH-LU factorization for \(n=32.768\).}
  \label{fig:dynrange}
\end{figure}

For the applications used for the numerical experiments in Section~\ref{sec:numexp}, the observed dynamic range during
\mcH-LU factorization, i.e., for all dense and lowrank data blocks for all intermediate and final data, is shown in
Figure~\ref{fig:dynrange}. As can be seen, it is much smaller compared to what is anticipated in the standard formats of
the IEEE-754 scheme. This permits to reduce the number of bits for the floating point storage not only for the mantissa
but also for the exponent part.

With \(e\) exponent bits, a dynamic range of \(\log_{10}(2^{2^e})\) is reached\footnote{This does not take into account
  differences between \emph{normalized} and \emph{denormalized} numbers in IEEE754.}. For a given data block
\(M \in \R^{n \times m}\) the dynamic range \(r := \log_{10} \frac{m_{\max}}{m_{\min}}\), with
\(m_{\max} := \max_{i,j} m_{ij}\) and \(m_{\min} := \min_{i,j} m_{ij}\), can therefore be represented by
\(e_r := \lceil \log_2 \log_2 r \rceil\) exponent bits. Together with the \(m_{\varepsilon}\) mantissa bits chosen based
on the required accuracy, we will use \(1-e_r-m_{\varepsilon}\) as the storage format.

\begin{remark}
  For easier conversion, all values are scaled by \(1/m_{\min}\) and shifted by \(1\), i.e., \(m_{ij} / m_{\min} + 1\),
  which ensures values bigger than 2 and as such only the lower \(e_r\) exponent bits need to be copied into compressed
  storage.
\end{remark}

This format is denoted \emph{AFL} and is similar to CPEN in \cite{AnzGruQui:2019}. However, it leads to a non-byte
aligned storage format as \(1 + e_r + m_{\varepsilon}\) is not necessarily a multiple of 8. The required bit handling may
increase the overhead while reading/writing data from/to memory.

As an alternative, \(m_{\varepsilon}\) is increased to \(m'_{\varepsilon}\) such that \(1 + e_r + m'_{\varepsilon} \mod
8 = 0\). The resulting format is denoted \emph{AFLP}.

For comparision, we will also use a format with an adaptive choice of mantissa bits but a constant number of 8 exponent
bits (\emph{BFL}), as this is identical to the single precision IEEE-754 format and an analog format with 11 exponent
bits (\emph{DFL}) which equals the corresponding double precision format. In both cases, it should be more efficient to
copy the exponent bits from the original (casted) values. Furthermore, to ensure byte-aligned storage, the mantissa bits
are again increased in both formats as with the AFLP format.

\begin{remark}
  The \emph{CPMS} format in \cite{AnzGruQui:2019} is comparable to DFL due to the adaptive mantissa choice and the 11
  exponent bits from the FP64 format, however the actual storage scheme is different due to separate mantissa and
  exponent storage.
\end{remark}

\begin{remark}
  An alternative to the IEEE-754 formats are \emph{posits} \cite{GusYon:2017}, which use \(n\) bits including \(es\)
  (maximal) exponent bits. This permits an adaptive storage (and compute) format depending on the accuracy \(\varepsilon\)
  \emph{and} the required dynamic range. However, standard CPUs and GPUs do not yet support this format.
\end{remark}

\subsection{Adaptive Precision Compression} \label{sec:adaprec}

For a matrix block \(M_{\clt,\cls}\) we assume a rank-\(k\) approximation \(U_{\clt,\cls} V_{\clt,\cls}^H\) with
\(\|M_{\clt,\cls} - U_{\clt,\cls} V_{\clt,\cls}^H \| \le \delta, \delta = \varepsilon ||M_{\clt,\cls}\)
(rf. \eqref{eqn:epsilon}).

With \(p\) different floating point formats with corresponding unit roundoffs \(u_i\), \(U_{\clt,\cls} V_{\clt,\cls}^H\)
is represented in \cite{AmeBoiBut:2022} as
\begin{align*}
  & U_{\clt,\cls} V_{\clt,\cls}^H = W \Sigma X^H := \\
  & \begin{pmatrix}
    W_0 & \ldots & W_{p-1}
  \end{pmatrix}
  \begin{pmatrix}
    \Sigma_0 & & \\
    & \ddots & \\
    & & \Sigma_{p-1}
  \end{pmatrix}
  \begin{pmatrix}
    X_0 & \ldots & X_{p-1}
  \end{pmatrix}^H
\end{align*}
with orthogonal \(W \in \R^{\#\clt \times k}, X \in \R^{\#\cls \times k}\) and \(\Sigma \in \R^{k \times k}\). The
subblocks \(W_i \in \R^{\#\clt \times k_i}, X_i \in \R^{\#\cls \times k_i}\) are then represented in the \(i\)'th
floating point format, while \(\Sigma_i \in \R^{k_i \times k_i}\) holds the singular values \(\sigma_j\) of the
corresponding subblock and is stored in the floating point format with highest accuracy (normally FP64). The
partitioning of the \(k = \sum_{i=0}^{p-1} k_i\) singular values is determined by the unit roundoffs \(u_i\) such that
\(\|\Sigma_i\| \approx \delta / u_i\), ensuring \cite[Theorem~2.1]{AmeBoiBut:2022} the overall error bound
\begin{displaymath}
  \|M_{\clt,\cls} - W \Sigma X^H \| \le \left( 2p - 1 + \sum_{i=1}^{p-1} \sqrt{r_i} u_i \right) \delta .
\end{displaymath}

This is based on a given set of floating point formats, which is normally defined by available hardware
support. However, using a general floating point compression scheme with adaptive error control, e.g., ZFP or AFL, one
may reverse the view and choose a precision \(\widetilde u_i\) such that \(\sigma_i \approx \delta / \widetilde
u_i\). With this, one can represent \(U_{\clt,\cls} V_{\clt,\cls}^H\) as
\begin{align*}
  & U_{\clt,\cls} V_{\clt,\cls}^H = W \Sigma X^H \approx \\
  & \begin{pmatrix}
    \widetilde w_0 & \ldots & \widetilde w_{k-1}
  \end{pmatrix}
  \begin{pmatrix}
    \sigma_0 & & \\
    & \ddots & \\
    & & \sigma_{k-1}
  \end{pmatrix}
  \begin{pmatrix}
    \widetilde x_0 & \ldots & \widetilde x_{k-1}
  \end{pmatrix}^H
\end{align*}
with \(\widetilde w_i\) (\(\widetilde x_i\)) being the \(i\)'th column of \(W\) (\(X\)), stored with precision
\(\widetilde u_i\).

The such extended lowrank compression shall be denoted \emph{A}daptive \emph{P}recision (vs. \emph{mixed precision})
compression for \emph{L}ow-\emph{R}ank matrices (APLR). The resulting storage format is denoted by \emph{ZFP-APLR} or \emph{AFL-APLR},
respectively. In an analog way, the schemes AFLP, BFL, DFL or any other floating point compressor can be extended.

%%%%%%%%%%%%%%%%%%%%%%%%%%%%%%%%%%%%%%%%%%%%%%%%%%%%%%%%%%%%%%%%%%%%%%%%%%%%%%%%%%%
%%%%%%%%%%%%%%%%%%%%%%%%%%%%%%%%%%%%%%%%%%%%%%%%%%%%%%%%%%%%%%%%%%%%%%%%%%%%%%%%%%%
%%%%%%%%%%%%%%%%%%%%%%%%%%%%%%%%%%%%%%%%%%%%%%%%%%%%%%%%%%%%%%%%%%%%%%%%%%%%%%%%%%%

\section{Compressed \mcH-arithmetic} \label{sec:zarith}

In \cite{AnzGruQui:2019}, the concept of a \emph{memory accessor} is introduced, which implements on-the-fly conversion
between the storage format and the computation format during arithmetic. In principle, the same could be applied to
\mcH-arithmetic as well. However, this would need a complete implementation of all basic linear algebra functions on
which \mcH-arithmetic is based, normally provided by vendor optimized libraries implementing the BLAS/LAPACK
\cite{lapack} function set.

As this would (most probably) result in a much less efficient implementation compared to existing BLAS/LAPACK libraries,
we refrained from this concept. Instead a \emph{semi-on-the-fly} approach was chosen for compressed \mcH-arithmetic,
where the conversion between the storage and the compute formats are performed \emph{before} and \emph{after}
the standard arithmetic functions. 

An example for this is presented in Algorithm~\ref{alg:zlrsvd} which is the compressed version of
Algorithm~\ref{alg:lrsvd}. At the beginning, the input data is decompressed from the storage format into the
computation format and at the end the output data is compressed back into the storage format.

\begin{algorithm}{Lowrank truncation with semi-on-the-fly data conversion}{alg:zlrsvd}
  \Procedure{truncate}{in: $U^c, V^c$, out: $W^c, X^c$}
  \State \([U,V] := \function{\bfseries decompress}[\(U^c,V^c\)]\);
  \State \([Q_U, R_U] := \function{qr}[\(U\)]\);
  \State \([Q_V, R_V] := \function{qr}[\(V\)]\);
  \State \([U_s,S_s,V_s] := \function{svd}[\(R_U \cdot R_V'\)]\);
  \State \(k := \function{rank}[ \(S_s\) ];\)
  \State \(W := Q_U \cdot  U_s(1:k,:) \cdot S_s(1:k,1:k) \);
  \State \(X := Q_V \cdot V_s(1:k,:)\);
  \State \([W^c,X^c] := \function{\bfseries compress}[\(W,X\)]\);
  \EndProcedure
\end{algorithm}

With this, any \mcH-arithmetic may simply be extended by the data conversion function calls and the data storage within
\mcH-matrices is replaced by the corresponding data storage format.

When using adaptive precision compression, e.g., AFL-APLR, the same concept may be applied reusing existing arithmetic
functions for \mcH-matrices. However, the orthogonality of the \(W\) and \(X\) factors need to be
ensured which may lead to additional computational costs. In principle, one may further optimize certain functions to
make use of the \(W \Sigma X^H\) representation, e.g., in Algorithm~\ref{alg:lrsvd} the multiplication with
\(S_s(1:k,1:k)\) can be omitted.

\subsection{\mcH-LU factorization}

When applying this concept for \mcH-LU factorization with standard arithmetic, i.e., as in Algorithm~\ref{alg:hmul} with
eager update application to destination blocks, the additional error during compression after each update has a
significant effect on the overall error of the factorization process, measured as \(|| I - A (LU)^{-1} ||_2\) with \(A\)
being the original \mcH-matrix and \(L, U\) its computed LU factors in compressed \mcH-matrix format. This is shown in
Figure~\ref{fig:luerror} (left):

\begin{figure}[htb]
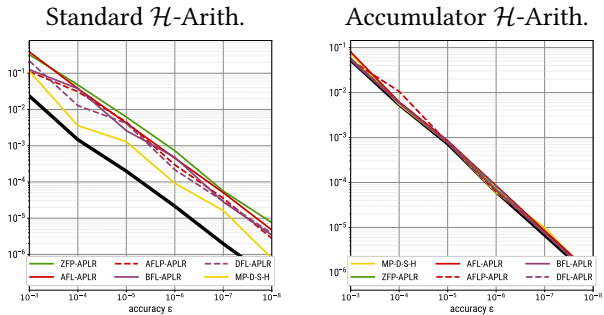

  \centering
  \pgfdeclareimage[width=.22\textwidth]{lapluerrstd}{data/lu/std/laplace-sphere-8/zerror-laplace-ap}
  \pgfdeclareimage[width=.22\textwidth]{lapluerraccu}{data/lu/laplace-sphere2-8/zerror-laplace-ap}
  \begin{tabular}{cc}
    Standard \mcH-Arith. & Accumulator \mcH-Arith. \\
    \pgfuseimage{lapluerrstd} & \pgfuseimage{lapluerraccu}
  \end{tabular}
  \caption{LU factorization error with standard (left) and accumulator based \mcH-arithmetic (right) for Laplace model
    problem \eqref{eqn:slp}.}
  \label{fig:luerror}
\end{figure}

To overcome this problem one could tighten the accuracy settings for compression during \mcH-arithmetic, which would
however deteriorate the compression ratio.

Alternatively, when using accumulator based \mcH-arithmetic the accumulator matrices do not need to be compressed as
this would have little effect on the overall memory usage (rf. Remark~\ref{rem:memaccu}). Therefore, no additional error
is introduces when computing the updates. The effect of this is visible in Figure~\ref{fig:luerror} (right), where the
error of the compressed arithmetic is identical to the uncompressed version. Since accumulator based \mcH-arithmetic is
also typically faster due to a reduced number of lowrank truncations, it is therefore favorable to use this arithmetic
when also using compression.

%%%%%%%%%%%%%%%%%%%%%%%%%%%%%%%%%%%%%%%%%%%%%%%%%%%%%%%%%%%%%%%%%%%%%%%%%%%%%%%%%%% 
%%%%%%%%%%%%%%%%%%%%%%%%%%%%%%%%%%%%%%%%%%%%%%%%%%%%%%%%%%%%%%%%%%%%%%%%%%%%%%%%%%%
%%%%%%%%%%%%%%%%%%%%%%%%%%%%%%%%%%%%%%%%%%%%%%%%%%%%%%%%%%%%%%%%%%%%%%%%%%%%%%%%%%%

\section{Numerical Experiments} \label{sec:numexp}

\subsection{Model Problems} \label{sec:numexpmp}

The first problem is based on a boundary element discretization for the Laplace single layer potential (Laplace SLP)
while the domain is defined by the unit sphere:
\begin{equation} \label{eqn:slp}
  \int_{\Omega} \frac{1}{\|x-y\|} u(x) dy = f(x), \quad x \in \Omega
\end{equation}
with \(\Omega = \set{ x \in \R^3 : \|x\|_2 = 1}\). Piecewise constant ansatz functions are used for the
discretization. Furthermore, standard admissibility
\begin{displaymath}
\min\left\{\op{diam}(t),\op{diam}(s)\right\} \le \eta \op{dist}(\clt,\cls)
\end{displaymath}
is applied for setting up the block tree.

The Matérn class of covariance functions forms the second model problem:
\begin{equation} \label{eqn:matcov}
  C(d, \sigma, \ell, \nu ) =
  \sigma^2 \frac{2^{1-\nu}}{\Gamma(\nu)}
  \left(\frac{\sqrt{2\nu}}{\ell}d\right)^\nu \mathcal{K}_\nu
  \left(\frac{\sqrt{2\nu}}{\ell}d\right),
\end{equation}
with \(d = \|x_i - x_j\|_2\) being the distance between two points \(x_i,x_j \in \Omega\) (randomly defined with unique
seed), \(\sigma^2 = 1\) the variance, \(\ell = 1\) a spatial range parameter and \(\nu = 1/3\) controlling the
smoothness of the random field. Furthermore, \emph{weak} admissibility is used as defined in \cite{HacKhoKri:2004},
which leads to fewer, larger lowrank blocks.

Both applications use a problem size of \(n=1.048.576\) and lowrank blocks are approximated via ACA~\cite{Beb:2000}. All
computations are performed in double precision (FP64) and this is also the default storage format, which serves as the
baseline in the following comparisons, indicated, if not otherwise noted, by a thick, black line. Results are therefore
presented \emph{relative} to the FP64 case.

\subsection{Machine and Software Settings} \label{sec:numexpmach}

\begin{figure*}[htb]
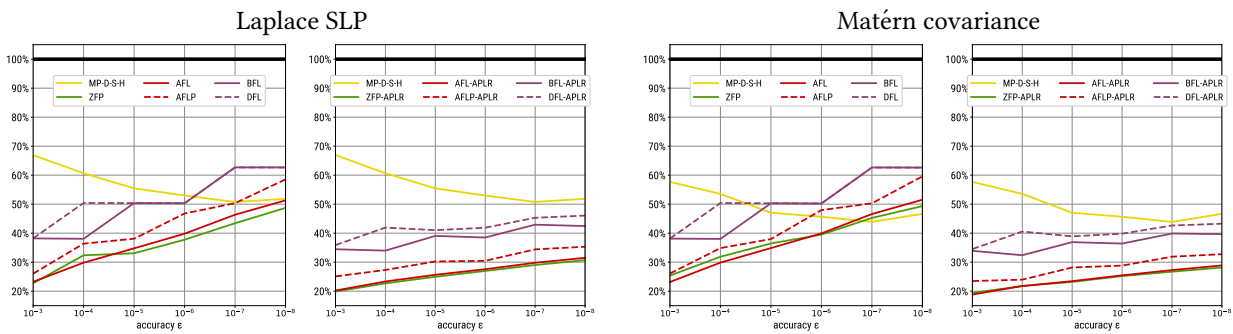

  \centering
  \pgfdeclareimage[width=.23\textwidth]{lapmem1}{data/mvm/laplace-sphere2-8/zmem-laplace-direct}
  \pgfdeclareimage[width=.23\textwidth]{matmem1}{data/mvm/matern-randsphere-1048576/zmem-materncov-direct}
  \pgfdeclareimage[width=.23\textwidth]{lapmem2}{data/mvm/laplace-sphere2-8/zmem-laplace-ap}
  \pgfdeclareimage[width=.23\textwidth]{matmem2}{data/mvm/matern-randsphere-1048576/zmem-materncov-ap}
  \begin{tabular}{cc}
    Laplace SLP & Matérn covariance \\
    \pgfuseimage{lapmem1}\pgfuseimage{lapmem2} & \pgfuseimage{matmem1}\pgfuseimage{matmem2}
  \end{tabular}
  \caption{Compression rates.}
  \label{fig:compmem}
\end{figure*}

All benchmarks are performed on a shared memory system with two AMD Epyc 9554 CPUs with 128 cores in total and 2x12 32GB
DDR5-4800 memory DIMMs.

For parallelization Intel TBB v2021.2 was used while Intel MKL v2022.0 provided the BLAS and LAPACK functions. Please
note, that here the sequential version was used as all parallelization is performed within the \mcH-arithmetic
itself. Furthermore, the AVX512 code path in MKL was activated. All code was compiled using GCC v12.1. Finally, ZFP~v1.0
was used.

The algorithms described in this work are implemented in the open source software HLR\footnote{\url{http://libhlr.org},
  programs: \emph{compress}, \emph{compress-lu}, \emph{mixedprec} and \emph{mixedprec-lu}}.

For all benchmarks the median of ten runs is presented.

\subsection{\mcH-compression} \label{sec:numexphcomp}

First, the compression of the \mcH-matrices after construction is examined. The compression is performed after fully
assembling the \mcH-matrices as these are also needed for the actual comparison. In practice this should be done
on-the-fly after each matrix block is constructed to reduce the overall memory consumption. The absolute memory usage of
the uncompressed \mcH-matrix varies between 16GB (\(\varepsilon = 10^{-3}\)) and 42GB (\(\varepsilon = 10^{-8}\)) for
the Laplace SLP problem and 8GB (\(\varepsilon = 10^{-3}\)) to 29 GB (\(\varepsilon = 10^{-8}\)) for the Matérn
covariance matrix.

In Figure~\ref{fig:compmem} the resulting compression rates, i.e., the ratios between the uncompressed and the
compressed \mcH-matrices, are shown for the compression schemes discussed in Section~\ref{sec:zhmat} and the mixed
precision approach from \cite{AmeBoiBut:2022} using FP64, FP32 and FP16 (denoted ``MP-D-S-H'').

Best compression is achieved by the ZFP library, however, only slightly more storage is needed with the AFL format. The
remaining storage schemes with rounded up mantissa bits (AFLP) or more exponent bits (BFL and DFL) follow. While for a
coarse accuracy the compression is significantly better than the mixed precision approach, the latter is on par (Laplace
SLP) or even exceeding (Matérn covariance) the other formats. The reason for this behaviour is that the mixed precision
format is able to use half precision floating point formats even for a high accuracy whereas the given error bound
applies to all data in a given floating point compression scheme like ZFP or AFL.

This changes with adaptive precision compression where the error bounds depend on the singular values. With this, the
corresponding compression rates can be significantly improved. Throughout the accuracy range much better results
compared to the mixed precision approach are achieved, reducing memory down to 20\% to about 30\% of the original FP64
storage with ZFP and AFL.

\begin{figure}[htb]
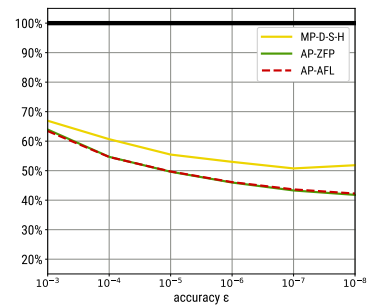

  \centering
  \pgfdeclareimage[width=.28\textwidth]{lapmemmp}{data/mvm/laplace-sphere2-8/zmem-laplace-mp}
  \pgfuseimage{lapmemmp}
  \caption{Compression rates with compression only applied to lowrank blocks for the Laplace model problem.}
  \label{fig:memmp}
\end{figure}

Part of the advantage of ZFP-APLR or AFL-APLR is, that in the mixed precision approach inadmissible blocks are not
compressed. Figure~\ref{fig:memmp} shows the results if only lowrank blocks are compressed. While ZFP and AFL are very
close together, both have, depending on the accuracy, a 5--10\% better compression rate compared to just using three
floating point types. It also shows the impact when applying compression to the dense matrices which still contribute
significantly to the overall storage even for rather large problem sizes.

\begin{figure*}[htb]
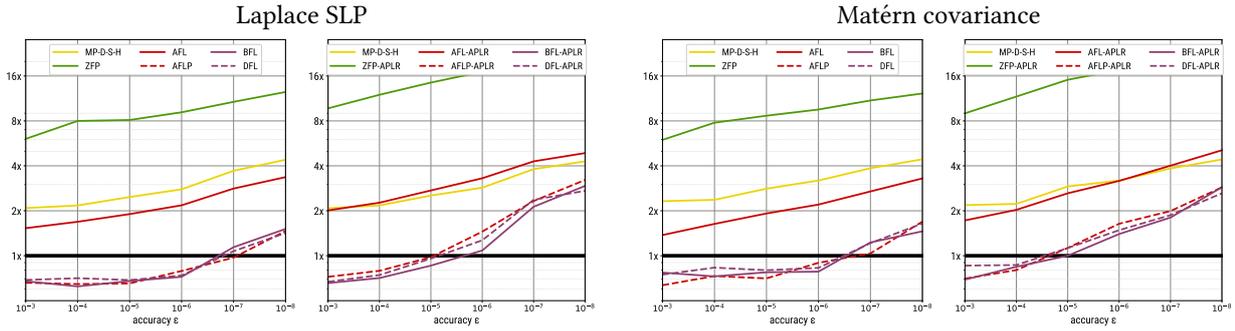

  \centering
  \pgfdeclareimage[width=.23\textwidth]{lapmvm1}{data/mvm/laplace-sphere2-8/zmvm-laplace-direct}
  \pgfdeclareimage[width=.23\textwidth]{matmvm1}{data/mvm/matern-randsphere-1048576/zmvm-materncov-direct}
  \pgfdeclareimage[width=.23\textwidth]{lapmvm2}{data/mvm/laplace-sphere2-8/zmvm-laplace-ap}
  \pgfdeclareimage[width=.23\textwidth]{matmvm2}{data/mvm/matern-randsphere-1048576/zmvm-materncov-ap}
  \begin{tabular}{cc}
    Laplace SLP & Matérn covariance \\
    \pgfuseimage{lapmvm1}\pgfuseimage{lapmvm2} & \pgfuseimage{matmvm1}\pgfuseimage{matmvm2}
  \end{tabular}
  \caption{Relative runtime of matrix vector multiplication.}
  \label{fig:compmvm}
\end{figure*}

\begin{figure}[htb]
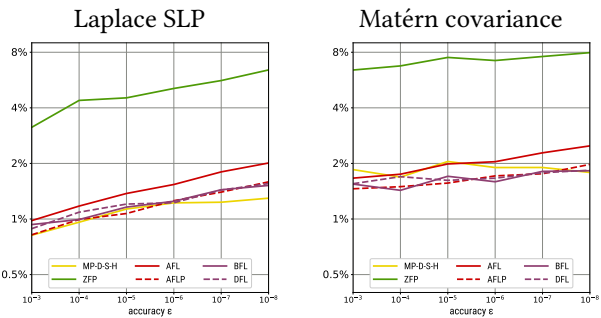

  \centering
  \pgfdeclareimage[width=.22\textwidth]{lapztime1}{data/mvm/laplace-sphere2-8/ztime-laplace-direct}
  \pgfdeclareimage[width=.22\textwidth]{matztime1}{data/mvm/matern-randsphere-1048576/ztime-materncov-direct}
  \begin{tabular}{cc}
    Laplace SLP & Matérn covariance \\
    \pgfuseimage{lapztime1} & \pgfuseimage{matztime1}
  \end{tabular}
  \caption{Compression time for \mcH-matrices.}
  \label{fig:comptime}
\end{figure}

Next, the compression time is shown in Figure~\ref{fig:comptime}. There, the runtime is presented relative to
the \mcH-matrix construction time. As can be seen, the additional compression overhead is negligible compared to setting
up the \mcH-matrix itself. However, this depends on the time to compute matrix coefficients or what lowrank
approximation method was used, which may be different for other applications.

\subsection{Matrix-Vector Multiplication} \label{sec:numexpmatvec}

Multiplication of a compressed \mcH-matrix with a vector will directly show the effect of the decompression speed of the
corresponding compression scheme with the results being presented in Figure~\ref{fig:compmvm}. Please note, that again
relative performance numbers compared to the uncompressed FP64 multiplication (indicated by the thick black baseline)
are used.

With compressed storage a better performance can be achieved for the lower accuracy regime using AFLP, BFL and
DFL. This indicates a memory bandwith limitation of \mcH-matrix vector multiplication for these problems. However, of
special importance for this is a \emph{fast} decompression, which does not hold for AFL and especially for
ZFP. However, for an accuracy towards \(\varepsilon = 10^{-8}\), the computational intensity of the matrix vector
multiplication increases due to a larger rank within the lowrank blocks, shifting the algorithm more into a compute
bound regime and thereby making the decompression overhead visible.

This can also be seen for the Matérn covariance problem, which typically shows a larger rank in the lowrank blocks
compared to the Laplace SLP model problem. Here, the break-even point for AFL, BFL and DFL is sooner and the overall
performance is worse compared to the original \mcH-matrix vector multiplication.

With additional computational overhead due to adaptive precision compression the performance gain is even more reduced
and requires a significantly higher runtime for the Matérn covariance model problem with a high accuracy, even surpassing
the mixed precision approach.

\begin{remark}
  It should be noted that with corresponding hardware and software support the mixed precision technique has the
  additional advantage of a faster execution time for those parts stored in single or half precision. This was not
  available for the benchmark system used in this work.
\end{remark}

\subsection{\mcH-LU Factorization} \label{sec:numexplu}

\begin{figure*}[htb]
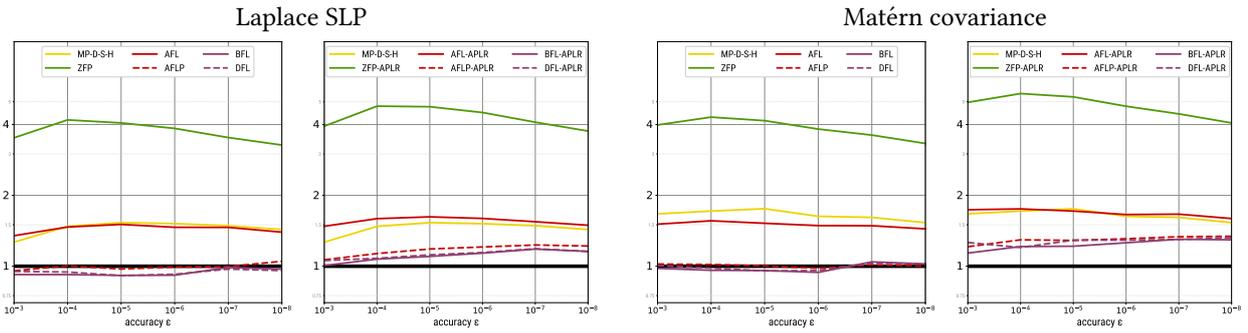

  \centering
  \pgfdeclareimage[width=.23\textwidth]{laplu1}{data/lu/laplace-sphere2-8/ztime-laplace-direct}
  \pgfdeclareimage[width=.23\textwidth]{matlu1}{data/lu/matern-randsphere-1048576/ztime-materncov-direct}
  \pgfdeclareimage[width=.23\textwidth]{laplu2}{data/lu/laplace-sphere2-8/ztime-laplace-ap}
  \pgfdeclareimage[width=.23\textwidth]{matlu2}{data/lu/matern-randsphere-1048576/ztime-materncov-ap}
  \begin{tabular}{cc}
    \multicolumn{1}{c}{Laplace SLP} & \multicolumn{1}{c}{Matérn covariance} \\
    \pgfuseimage{laplu1} \pgfuseimage{laplu2} & \pgfuseimage{matlu1} \pgfuseimage{matlu2}
  \end{tabular}
  \caption{Relative runtime of \mcH-LU factorization.}
  \label{fig:complu}
\end{figure*}

In Figure~\ref{fig:complu} the relative runtimes of the \mcH-LU factorization for the model problems with and without
adaptive precision compression are shown.

Unexpectedly, the runtime with BFL and DFL is slightly faster than the uncompressed version for the Laplace SLP problem,
indicating some memory bandwidth influence on the performance of the \mcH-LU factorization. The slightly costlier AFLP
shows the same performance as the uncompressed \mcH-LU as is also the case for BFL/DFL for the Matérn covariance
matrix.

About 50\% slower runtimes are achieved with AFL or mixed precision storage. ZFP again shows a much worse performance
and is about four times slower compared to the uncompressed version.

With adaptive precision compression the picture is very similar though with slightly increased runtimes. The most
notable difference is, that the performance advantage of BFL/DFL or the on par performance of AFLP turns into a slight
performance disadvantage compared to the uncompressed \mcH-LU. 

\begin{figure*}[htb]
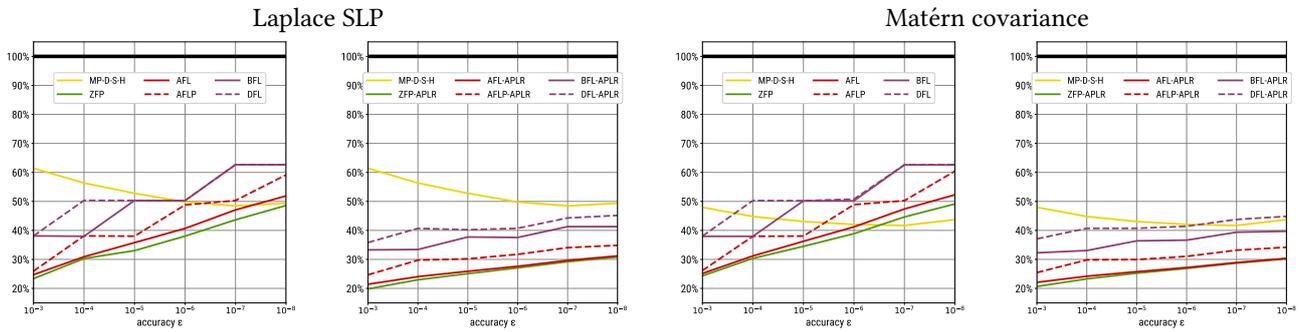

  \centering
  \pgfdeclareimage[width=.23\textwidth]{laplumem1}{data/lu/laplace-sphere2-8/zmem-laplace-direct}
  \pgfdeclareimage[width=.23\textwidth]{matlumem1}{data/lu/matern-randsphere-1048576/zmem-materncov-direct}
  \pgfdeclareimage[width=.23\textwidth]{laplumem2}{data/lu/laplace-sphere2-8/zmem-laplace-ap}
  \pgfdeclareimage[width=.23\textwidth]{matlumem2}{data/lu/matern-randsphere-1048576/zmem-materncov-ap}
  \begin{tabular}{cccc}
    \multicolumn{2}{c}{Laplace SLP} & \multicolumn{2}{c}{Matérn covariance} \\
    \pgfuseimage{laplumem1} & \pgfuseimage{laplumem2} & \pgfuseimage{matlumem1} & \pgfuseimage{matlumem2}
  \end{tabular}
  \caption{Compression rates for \mcH-LU factors.}
  \label{fig:complumem}
\end{figure*}

However, the overhead is still small compared to the memory savings, which are similar to the original \mcH-matrix as
can be seen in Figure~\ref{fig:complumem}. In absolute terms the memory footprint of the uncompressed factors is slightly
larger compared to the \mcH-matrix, i.e., 18GB (\(\varepsilon = 10^{-3}\)) to 48GB (\(\varepsilon = 10^{-8}\)) for the
Laplace SLP problem and 12GB (\(\varepsilon = 10^{-3}\)) to 40 GB (\(\varepsilon = 10^{-8}\)) for the Matérn covariance
problem.

%%%%%%%%%%%%%%%%%%%%%%%%%%%%%%%%%%%%%%%%%%%%%%%%%%%%%%%%%%%%%%%%%%%%%%%%%%%%%%%%%%% 
%%%%%%%%%%%%%%%%%%%%%%%%%%%%%%%%%%%%%%%%%%%%%%%%%%%%%%%%%%%%%%%%%%%%%%%%%%%%%%%%%%%
%%%%%%%%%%%%%%%%%%%%%%%%%%%%%%%%%%%%%%%%%%%%%%%%%%%%%%%%%%%%%%%%%%%%%%%%%%%%%%%%%%%

\section{Conclusion} \label{sec:conclude}

The application of binary compression to the data blocks in \mcH-matrices effectively decreases the memory costs further
and shows a significant advantage to other available formats, e.g., mixed precision schemes, especially if adaptive
precision compression for lowrank data is used.

Furthermore, using such additional compression can be done with little effect on the parallel performance of typical
\mcH-arithmetic operations, sometimes even resulting in a performance advantage due to decreased memory bandwidth
utilization.

Essential to this is a fast compression scheme with adaptive error control as is available with ZFP or the presented
IEEE754 based formats. It would be of high interest to look into other compressors in the future in the hope that
compression rates or arithmetic performance can be further improved.

%%% Local Variables: 
%%% mode: latex
%%% TeX-master: "article"
%%% ispell-local-dictionary: "american"
%%% fill-column: 120
%%% End: 

%%%%%%%%%%%%%%%%%%%%%%%%%%%%%%%%%%%%%%%%%%%%%%%%%%%%%%%%%%%%%%%%%%%%%%%%%%%%%%%%%%%%%%%%%
%%%%%%%%%%%%%%%%%%%%%%%%%%%%%%%%%%%%%%%%%%%%%%%%%%%%%%%%%%%%%%%%%%%%%%%%%%%%%%%%%%%%%%%%%

\printbibliography

\end{document}